\documentclass[preprint,aps,showpacs]{revtex4}
\usepackage{graphicx,amsfonts}
\usepackage{epsfig}
\usepackage{dcolumn}
\usepackage{bm}
\usepackage[active]{srcltx}
\begin{document}

\title{ 
\vglue -0.5cm
 \hfill{\small IFT-P.027/2004} \\
\hfill{\small June 2004}\\
\vglue 0.5cm
Supersymmetric 3-3-1 model with right-handed
neutrinos}
\author{J. C. Montero$^a$\footnote{E-mail address: 
montero@ift.unesp.br}, 
V. Pleitez~$^a$\footnote{E-mail address: vicente@ift.unesp.br} 
and M. C. Rodriguez$^b$\footnote{E-mail address: mcrodriguez@fisica.furg.br}
}
\affiliation{
$^a$ Instituto de F\'\i sica Te\'orica\\
Universidade Estadual Paulista\\
Rua Pamplona, 145\\ 
01405-900-- S\~ao Paulo, SP\\
Brazil\\
$^b$ Funda\c c\~ao Universidade Federal do Rio Grande/FURG \\
Departamento de F\'\i sica\\
Av. It\'alia, km 8, Campus Carreiros \\
96201-900, Rio Grande, RS\\
 Brazil }
\date{\today}

\begin{abstract}
We consider the supersymmetric extension of the 3-3-1 model with right-handed
neutrinos. We study the mass spectra in the scalar and pseudoscalar sectors, and
for a given set of the input parameters, we find that the lightest scalar in the
model has a mass of 130 GeV and the lightest pseudoscalar has mass of 5
GeV. However, this pseudoscalar decouples from the $Z^0$ at high energy scales
since it is almost a singlet under $SU(2)_L\otimes U(1)_Y$.
\end{abstract}

\maketitle


\section{Introduction}
\label{sec:intro}
 
Models with $SU(3)_C\otimes SU(3)_L\otimes U(1)_N$ gauge symmetry (called 3-3-1
models for short) are interesting  
possibilities for the physics at the TeV scale~\cite{331,outros,mpp}. 
At low energies they coincide with the standard model and some of them give at
least partial explanation to some fundamental questions that are accommodated
but not explained by the standard model. For instance, in order to
cancel the triangle anomalies, together with asymptotic freedom in QCD, 
the model predicts that the number of generations must be three and only three;
{\it (ii)} the model of Ref.~\cite{331} predicts that
$(g'/g)^2=\sin^2\theta_W/(1-4\sin^2\theta_W)$, thus there is a Landau pole at
the energy scale $\mu$ at which $\sin^2\theta_W(\mu)=1/4$.  According to
recent calculations $\mu\sim4$ TeV~\cite{phf,pl331} ; 
{\it (iii)} the quantization of the electric charge~\cite{pr} and the vectorial
cha\-rac\-ter of the electromagnetic interactions~\cite{cp} do not depend on the
nature of the neutrinos i.e., if they are Dirac or Majorana particles; {\it
(iv)} as a consequence of item ii) above, the model possesses perturbative
${\cal N}=1$ supersymmetry naturally at the $\mu$ scale~\cite{331susy,marcos};
{\it (v)} the Peccei-Quinn~\cite{pq} symmetry occurs naturally in these
models~\cite{axion331}; {\it (vi)} since one generation of quarks is treated
differently from the others this may be lead to a natural explanation for  the
large mass of the top quarks~\cite{longvan}. Moreover, if right-handed neutrinos
are considered transforming non-trivially~\cite{mpp}, 3-3-1
models~\cite{331,outros} can be embedded in a model with 3-4-1 gauge symmetry in
which leptons transform as $(\nu_l,l,\nu^c_l,\,l^c)_L\sim({\bf1},{\bf4},0)$
under each gauge factors~\cite{su4}. The $SU(3)_L$ symmetry is
possibly the largest symmetry  involving the known leptons (and $SU(4)_L$ if
right-handed neutrinos do really exist). This make 3-3-1 or 3-4-1 models
interesting by their own.      
 
Models with  $SU(3)$ (or $SU(4)$) electroweak symmetry may have doubly charged 
vector bosons. These sort of bileptons may be detected by 
measuring several left-right asymmetries in M\o ller scattering~\cite{assi} or,
in future lepton-lepton accelerators. It is interesting that one these parity
non-conserving asymmetry was observed in M\o ller scattering for the
first time only recently and its value is, at present, in agreement with the
standard model prediction~\cite{e158b}. However, in the future   
those asymmetries in $e^-e^-$ colliders~\cite{assi} may also be used
for searching doubly charged particles and a heavy neutral
$Z^{\prime 0}$ vector bosons, which is also a prediction of those models,  may
be discovered in $e^-\mu^-$ colliders~\cite{emu}. Singly and
doubly charged vector bileptons may also be produced in
$e^-\gamma$~\cite{egamma} or 
$\gamma\gamma$~\cite{gammagamma} or hadron~\cite{dion} colliders.  
New heavy quarks are also part of the electroweak quark multiplets in the
mi\-ni\-mal model representation. They are singlets under the  standard model
$SU(2)_L\otimes U(1)_Y$ group symmetry and in some versions of the 3-3-1 models
the electric charges of these heavy quarks are different from the usual ones, so
that it can be used to distinguish such models from 
their viable competitors. In fact, 
the $p\overline{p}$ production of these exotic quarks at the energies 
of the Tevatron have been studied in Ref.~\cite{djm} where a lower bound
of 250 GeV on their masses was found.
This sort of models are also predictive with respect to neutrino 
masses~\cite{numass}: they can implement the large mixing angle 
MSW solution to the solar neutrino issue~\cite{lma}, and also the almost 
bi-maximal mixing matrix in the lepton sector~\cite{pmns}. 

Gauge models  based on the 3-3-1 gauge symmetry
can have several representation contents~\cite{331,mpp} depending on the
embedding of the charge operator in the $SU(3)_{L}$ generators,
\begin{equation}
\frac{Q}{e}= \frac{1}{2}(\lambda_3- \vartheta \lambda_8)+N \,\ I,
\label{co}
\end{equation}
where the $\vartheta$ parameter defines two different representation contents, 
$N$ denotes the $U(1)_{N}$ charge and $\lambda_3$, $\lambda_8$ are the diagonal 
generators of $SU(3)$. 
The supersymmetric version of the model of Ref.~\cite{331}, 
with $\vartheta=\sqrt3$, 
has already been considered in Refs.~\cite{331susy,marcos}. In this work we
build the supersymmetric model based on the representation content of
the 3-3-1 model of Refs.~\cite{mpp} which corresponds the the case when
$\vartheta=1/\sqrt3$. 

The outline of this paper is as follows. In
Sec.~\ref{sec:model} we review the non-supersymmetric 3-3-1 model with
right-handed neutrinos, introduce the respective superpartners and the
superfields. The Lagrangian of this 3-3-1 supersymmetric model, including the
soft term, is considered in Sec.~\ref{sec:lagrangian}; the scalar potential and
the scalar mass spectra are given in Sec.~\ref{sec:potential}. Finally, the last
section is devoted to our conclusions.

\section{The model}
\label{sec:model}

In this section (Sec.~\ref{subsec:nonsusy}) we review the non-supersymmetric
3-3-1 model with right-handed neutrinos of Refs.~\cite{mpp} and add the 
superpartners (Sec.~\ref{subsec:susyp}). The superfields are introduced in
Sec.~\ref{subsec:superfields}. 
 
\subsection{The representation content}
\label{subsec:nonsusy}

Let us first summarize the non-supersymmetric model with the 
charge operator defined by Eq.~(\ref{co}) with $\vartheta=1/\sqrt3$~\cite{mpp},
i.e.,.
\begin{eqnarray}
\frac{Q}{e} &=& \frac{1}{2}\left(\lambda_3-\frac{1}{ \sqrt{3}}\lambda_8\right)
+N {\bf I},
\label{compp}
\end{eqnarray}
which implies leptons transforming under the 3-3-1 factors as 
\begin{equation} 
L_{aL} =
\left( \begin{array}{c} \nu_a \\ 
                        l_a \\
                        \nu^{c}_{a}          
\end{array} \right)_{L} \sim ( {\bf 1},{\bf 3},-1/3), 
 \label{trip}
\end{equation}
with $a= e, \mu, \tau $ and $\nu_{a}^{c}=C \bar{ \nu_{a}}^{T}$, plus the
singlets 
\begin{eqnarray}
l^{c}_{aL} \sim ({\bf 1},{\bf 1},1).
\label{singletos}
\end{eqnarray}

In the quark sector we have the first two families transforming as 
antitriplets of
$SU(3)_L$  
\begin{equation} 
Q_{\alpha L} = 
\left( \begin{array}{c} d_\alpha \\ 
                        u_\alpha \\
                        d^{ \prime}_\alpha \end{array} \right)_{L} \sim 
\left( {\bf 3},{\bf 3^{*}},0 \right), \;\;\alpha=1,2;
\label{quark323} 
\end{equation}
with the respective singlets 
\begin{equation}
u^{c}_{\alpha L} \sim 
\left( {\bf 3^{*}},{\bf 1},-2/3 \right), \;\;
d^{c}_{\alpha L},\;
d^{ \prime c}_{\alpha L}  \sim 
\left( {\bf 3^{*}},{\bf 1},1/3 \right). 
\label{sing1}
\end{equation}

The third family transforms as triplet under $SU(3)_L$ in such a way that,
the triangle anomaly cancels out only among the three
families and taken into account also the color charges. 
\begin{eqnarray} 
Q_{3L} &=& 
\left( \begin{array}{c} u_3 \\ 
                        d_3 \\
                        u^{ \prime}          
\end{array} \right)_{L} \sim ( {\bf 3},{\bf 3},1/3),
\label{quarks3} 
\end{eqnarray}
and their respective singlets  
\begin{equation}
u^{c}_{3L},\; u^{ \prime c}_{L} \sim ( {\bf 3^{*}},{\bf 1},-2/3), \;\;
d^{c}_{3L} \sim ( {\bf 3^{*}},{\bf 1},1/3 ).
\label{qsingletos}
\end{equation}

In the scalar sector only two triplets $\eta\sim({\bf1}, {\bf3}, -1/3)$
and $\rho\sim({\bf1},{\bf3}, 2/3)$ are necessary to break appropriately the
gauge symmetry and also to give the correct mass to all the fermions in the
model. However, to eliminate flavor changing neutral currents we add an extra
scalar triplet transforming like $\eta$.

\begin{eqnarray} 
\eta &=& 
\left( \begin{array}{c} \eta^{0}_{1} \\ 
                        \eta^{-} \\
                        \eta^{0}_{2} 
\end{array} \right), \;\; 
\chi = 
\left( \begin{array}{c}  \chi^{0}_{1} \\ 
                        \chi^{-} \\
                        \chi^{0}_{2} 
\end{array} \right) \sim ( {\bf 1},{\bf 3},-1/3 ), 
\nonumber \\
\rho &=& 
\left( \begin{array}{c}  \rho^{+}_{1} \\ 
                        \rho^{0} \\
                        \rho^{+}_{2} 
\end{array} \right) \sim ( {\bf 1},{\bf 3},2/3 ),  
\label{esctrip}
\end{eqnarray}
and we will denote the vacuum expectation values which are different from zero 
as $v=\langle\eta^0_1\rangle/\sqrt2$, $w=\langle\chi^0_2\rangle/\sqrt2$ and
$u=\langle\rho^0\rangle/\sqrt2$.

\subsection{Supersymmetric partners}
\label{subsec:susyp}

Here we will follow the usual notation writing for a given fermion $f$, the
respective sfermions by $\tilde{f}$ {\it i.e.}, $\tilde{l}$ and $\tilde{q}$
denote sleptons and squarks respectively. Then, we have the following additional
representations

\begin{eqnarray} 
\tilde{Q}_{\alpha L}& =& 
\left( \begin{array}{c} \tilde{d}_\alpha \\ 
                        \tilde{u}_\alpha \\
                        \tilde{d}^{ \prime}_\alpha          
			\end{array}\right)_L 
\sim ( {\bf 3},{\bf 3^{*}},0 ), \;
\tilde{Q}_{3L} =
\left( \begin{array}{c} \tilde{u}_3 \\ 
                        \tilde{d}_3 \\
                        \tilde{u}^{ \prime}       
			\end{array} 
\right)_{L}\!\!\!\! \sim ( {\bf 3},{\bf 3},1/3 ), \nonumber \\
\tilde{L}_{aL} &=& 
\left( \begin{array}{c} \tilde{ \nu}_{a} \\ 
                        \tilde{l_a} \\
                        \tilde{ \nu}^{c}_{a}          
\end{array} \right)_{L} \sim ( {\bf 1},{\bf 3},-1/3 ), 
\label{susytri}
\end{eqnarray}

\begin{eqnarray}
\lefteqn{\tilde{l}^{c}_{aL}\sim ({\bf 1},{\bf 1},1),}
\nonumber \\ &&
\tilde{u}^{c}_{i L},\,\tilde{u}^{ \prime c}_{L} 
\sim ( {\bf 3}^*,{\bf 1},-2/3),\; 
\tilde{d}^{c}_{iL}, \tilde{d}^{ \prime c}_{\alpha L}
 \sim ( {\bf 3}^*,{\bf 1},1/3),
\label{ss}
\end{eqnarray}
with $a=e, \mu , \tau$; $i=1,2,3$; and $\alpha=1,2$. 
However, when considering quark (or squark) singlets of a given charge we will
use the notation $u^c_{iL},d^c_{iL}$ ($\tilde{u}_{iL},\tilde{d}^c_{iL}$ with
$i(j)=1,2,3$.  

The supersymmetric partner of the scalar Higgs fields, the higgsinos, are
\begin{eqnarray} 
\tilde{ \eta} &=& 
\left( \begin{array}{c} \tilde{ \eta}^{0}_{1} \\ 
                        \tilde{ \eta}^{-} \\
                        \tilde{ \eta}^{0}_{2} 
\end{array} \right),\; 
\tilde{ \chi} = 
\left( \begin{array}{c} \tilde{ \chi}^{0}_{1} \\ 
                        \tilde{ \chi}^{-} \\
                        \tilde{ \chi}^{0}_{2} 
\end{array} \right) \sim ( {\bf 1},{\bf 3},-1/3 ), 
\nonumber
\\
\tilde{ \rho} &=& 
\left( \begin{array}{c} \tilde{ \rho}^{+}_{1} \\ 
                        \tilde{ \rho}^{0} \\
                        \tilde{ \rho}^{+}_{2} 
\end{array} \right) \sim ( {\bf 1},{\bf 3},2/3 ),
\label{esca} 
\end{eqnarray}
and the respective extra higgsinos, needed to cancel the chiral anomaly of the
higgsinos in Eq.~(\ref{esca}), are
\begin{eqnarray} 
\tilde{\eta}^{\prime} &=& 
\left( \begin{array}{c} \tilde{\eta}^{\prime 0}_{1} \\ 
                        \tilde{\eta}^{\prime +} \\
                        \tilde{\eta}^{\prime 0}_{2}          
			\end{array} \right),
\tilde{\chi}^{\prime} = 
\left( \begin{array}{c} \tilde{ \chi}^{\prime 0}_{1} \\ 
                        \tilde{ \chi}^{\prime +} \\
                        \tilde{ \chi}^{\prime 0}_{2} 
\end{array} \right) \sim ( {\bf 1},{\bf 3^{*}},1/3 ), 
\nonumber \\
\tilde{\rho}^{\prime} &=& 
\left( \begin{array}{c} \tilde{\rho}^{\prime -}_{1} \\ 
                        \tilde{\rho}^{\prime 0} \\
                        \tilde{\rho}^{\prime -}_{2} 
\end{array} \right) \sim ( {\bf 1},{\bf 3^{*}},-2/3 ), 
\label{escac}
\end{eqnarray}
and the corresponding scalar partners denoted by $\eta^{\prime}$,$\chi^{\prime}$,
$\rho^{\prime}$, with the same charge assignment as in Eq.~(\ref{escac}), 
and with the following VEVs: 
$v^{\prime}=\langle \eta^{\prime 0}_1 \rangle/\sqrt2$,
$w^{\prime}=\langle \chi^{\prime 0}_2\rangle /\sqrt2$ and  
$ u^{\prime}=\langle \rho^{\prime 0} \rangle/\sqrt2$. 
This complete the representation content of this supersymmetric model.  

Concerning the gauge
bosons and their superpartners, if we denote the gluons by $g^b$ the respective
superparticles, the gluinos, are denoted by $\lambda^b_{C}$, with 
$b=1, \ldots,8$; and in the electroweak sector we have
$V^b$, the gauge boson of $SU(3)_{L}$, and their gauginos partners  
$\lambda^b_{A}$; finally we have the gauge boson of 
$U(1)_{N}$, denoted by $V^\prime$, and its supersymmetric partner $\lambda_{B}$.
This is the total number of fields in the minimal supersymmetric extension of
the 3-3-1 model of Refs.~\cite{mpp}. 

\subsection{Superfields}
\label{subsec:superfields}

The superfields formalism is useful in writing the Lagrangian which is 
manifestly invariant under the supersymmetric transformations~\cite{wb} with 
fermions and scalars put in chiral superfields while the gauge bosons in 
vector superfields. As usual the superfield of a field $\phi$ will be denoted
by $\hat{\phi}$~\cite{mssm}.
The chiral superfield of a multiplet $\phi$ is denoted by 
\begin{eqnarray}
\hat{\phi}\equiv\hat{\phi}(x,\theta,\bar{\theta})&=& \tilde{\phi}(x) 
+ i \; \theta \sigma^{m} \bar{ \theta} \; \partial_{m} \tilde{\phi}(x) 
+\frac{1}{4} \; \theta \theta \; \bar{ \theta}\bar{ \theta} \; \Box 
\tilde{\phi}(x) \nonumber \\ 
& & \mbox{} +  \sqrt{2} \; \theta \phi(x) 
+ \frac{i}{ \sqrt{2}} \; \theta \theta \; \bar{ \theta} \bar{ \sigma}^{m}
\partial_{m}\phi(x)                   
\nonumber \\ && \mbox{}+  \theta \theta \; F_{\phi}(x), 
\label{phi}
\end{eqnarray}
while the vector superfield is given by
\begin{eqnarray}
\hat{V}(x,\theta,\bar\theta)&=&-\theta\sigma^m\bar\theta V_m(x)
+i\theta\theta\bar\theta
\overline{\tilde{V}}(x)-i\bar\theta\bar\theta\theta
\tilde{V}(x)\nonumber \\ &+&\frac{1}{2}\theta\theta\bar\theta\bar\theta D(x).
\label{vector}
\end{eqnarray}
The fields $F$ and $D$ are auxiliary fields which are needed to
close the supersymmetric algebra and eventually will be eliminated using 
their motion equations. 

For fermion superfields we use the notation
\begin{equation}
\hat{L}_{aL},\; \hat{l}^{c}_{aL},\;   
\hat{Q}_{\alpha L},\;  \hat{Q}_{3L},\;
\hat{u}^{c}_{iL},\;
\hat{d}^{c}_{iL},\;  
\hat{u}^{\prime c}_{L},\; \hat{d}^{\prime c}_{\alpha L}. 
\label{superfermions}
\end{equation}

For scalar superfields we write: $\hat{ \eta}, \;
\hat{ \chi},\;\hat{ \rho}$ and similar expressions for $\hat{\eta}^{\prime}$,
$\hat{\chi}^{\prime}$, $\hat{\rho}^{\prime}$ and we must change  
$\mbox{(field)}$ by $\mbox{(field)}^{\prime}$.

The vector superfield for the gauge bosons of each factor $SU(3)_C$, $SU(3)_L$
and $U(1)_N$ are denoted by $\hat{V}_C,\hat{\bar V}_C$;
$\hat{V},\hat{\bar V}$; and $\hat{V^\prime}$,
respectively, where we have defined $\hat{V}_C=T^b\hat{V}^b_C$, 
$\hat{V}=T^b\hat{V}^b$; $\hat{\bar V}_C=\bar{T}^b\hat{V^b}_C$,
$\hat{\bar V}=\bar{T}^b\hat{V^b}$;
$T^b=\lambda^b/2$, $\bar{T}^b=-\lambda^{*b}/2$ are the generators of triplet
and antitriplets representations, respectively, and $\lambda^b$ are the
Gell-Mann matrices.   
 
\section{The Lagrangian}
\label{sec:lagrangian}

The Lagrangian of the model has the following form 
\begin{eqnarray} 
   {\cal L}_{331S} &=& {\cal L}_{SUSY} + {\cal L}_{\mbox{soft}}, 
\label{lagra}
\end{eqnarray}
where ${\cal L}_{SUSY}$ is the supersymmetric part and  
${\cal L}_{\mbox{soft}}$ the soft terms breaking explicitly the supersymmetry.

\subsection{The supersymmetric Lagrangian}
\label{subsec:susy}

The supersymmetric part of the Lagrangian is decomposed  in the
lepton, quark, gauge, and the scalar sectors as follow:
\begin{eqnarray} 
{\cal L}_{SUSY} &=&   {\cal L}_{\mbox{Lepton}}+ {\cal L}_{\mbox{Quark}}+ 
{\cal L}_{\mbox{Gauge}}+ {\cal L}_{\mbox{Scalar}}, 
\label{susyterm}
\end{eqnarray}
where
\begin{eqnarray}
{\cal L}_{\mbox{Leptons}} 
&=& \int\!\! d^{4}\theta\left[\,\hat{
\bar{L}}_{aL}e^{2g\hat{V}-\frac{g'}{3}\hat{V}'}  
\hat{L}_{aL}+\hat{\bar{l}}^c_{aL}e^{g^\prime \hat{V}^\prime}
\hat{l}^c_{aL}\right], 
\label{lagrangiana1}
\end{eqnarray}
in the lepton sector, we have omitted the sum over the three lepton family for
simplicity, and
\begin{eqnarray}
{\cal L}_{\mbox{Quarks}} 
&=& \int\!\! d^{4}\theta\!\!\left[
\hat{\bar{Q}}_{\alpha L}
e^{[2(g_s\hat{V}_{C}+g\hat{\bar{V}})]} \hat{Q}_{\alpha L}+\hat{\bar{Q}}_{3L}
e^{[2(g_s\hat{V}_{C}+g\hat{V})+\frac{g'}{3}\hat{V}']} \hat{Q}_{3L}  
 \,\right. 
\nonumber \\&+& 
\left.\,\hat{ \bar{u}}^{c}_{iL}
e^{[2g_s \hat{\bar{V}}_C-\frac{2g'}{3}\hat{V}']} \hat{u}^{c}_{iL} 
+\hat{ \bar{d}^c}_{iL}
e^{[2g_s \hat{ \bar{V}}_{C}+\frac{g'}{3}\hat{V}']} \hat{d}^{c}_{iL}\right. 
\nonumber \\&+& 
\left.\,\hat{\bar{u}}^{\prime c}_L
e^{[2g_s \hat{ \bar{V}}_{C}-\frac{2g'}{3}\hat{V}']} \hat{u}^{\prime c}_L 
+\hat{ \bar{d}^c}^{\prime }_{\alpha L}
e^{[2g_s \hat{ \bar{V}}_{C}+\frac{g'}{3}\hat{V}']} 
\hat{d}^{\prime c}_{\alpha L} \right], 
\label{lagrangiana2}
\end{eqnarray}
in the quark sector, and we have denoted $g_s,g,g^\prime$ the 
gauge coupling constants for the $SU(3)_C,SU(3)_L,U(1)_N$ factors, respectively.
In the gauge sector we have
\begin{eqnarray}
{\cal L}_{\mbox{Gauge}} 
&=&  \frac{1}{4} \int  d^{2}\theta\; 
\left[ \cal {W}_{C}\cal{W}_{C}+{\cal W}\cal{W}+{\cal W}^{ \prime}{\cal W}^{
\prime}\,  \right] \nonumber \\
&+& \frac{1}{4} \int  d^{2}\bar{\theta}\; 
\left[ \bar{\cal W}_{C}\bar{\cal W}_{C}+\bar{\cal W}\bar{\cal W}+  
\bar{\cal W}^{ \prime}\bar{\cal W}^{ \prime}\, 
\right],
\label{lagrangiana3}
\end{eqnarray}
where $\cal{W}_{C}$, $\cal{W}$ e $\cal{W}^{ \prime}$ are fields that can be
written as follows 
\begin{eqnarray}
\cal{W}_{\zeta C}&=&- \frac{1}{8g_s} \bar{D} \bar{D} e^{-2g_s \hat{V}_{C}} 
D_{\zeta} e^{2g_s \hat{V}_{C}},\nonumber \\ 
\cal{W}_{\zeta}&=&- \frac{1}{8g} \bar{D} \bar{D} e^{-2g \hat{V}} 
D_{\zeta} e^{2g \hat{V}}, \nonumber \\
\cal{W}^{\prime}_{\zeta}&=&- \frac{1}{4} \bar{D} \bar{D} D_{\zeta} 
\hat{V}^{\prime}, \,\ \zeta=1,2.
\label{cforca}
\end{eqnarray}

Finally, in the scalar sector we have
\begin{eqnarray}
{\cal L}_{\mbox{Escalar}} 
&=& \int d^{4}\theta\;\left[\,
\hat{ \bar{ \eta}}\,e^{[2g\hat{V}-\frac{g'}{3}\hat{V}']} \hat{ \eta} +
\hat{ \bar{ \chi}}\,e^{[2g\hat{V}-\frac{g'}{3}\hat{V}']} \hat{ \chi} + 
\hat{ \bar{ \rho}}\,e^{[2g\hat{V}+\frac{2g'}{3}\hat{V}']} \hat{ \rho} 
\right. \nonumber \\
&+& \left.\,\hat{ \bar{ \eta}}^{\prime}\,e^{[2g\hat{ \bar{V}}+
\frac{g'}{3}\hat{V}']} \hat{ \eta}^{\prime} + 
\hat{ \bar{ \chi}}^{\prime}\,e^{[2g\hat{ \bar{V}}+
\frac{g'}{3}\hat{V}']} \hat{ \chi}^{\prime} +
\hat{ \bar{ \rho}}^{\prime}\,e^{[2g\hat{ \bar{V}}-
\frac{2g'}{3}\hat{V}']} \hat{ \rho}^{\prime} \right] \nonumber \\
&+& \int d^{2}\theta W+ \int d^{2}\bar{ \theta}\bar{W} \!,\hspace{2mm} 
\label{escalagra}
\end{eqnarray}
where $W$ is the superpotential. 

\subsection{Superpotential}
\label{subsec:sp}

The superpotential of the model is decomposed as follows
\begin{eqnarray}
W&=&\frac{W_{2}}{2}+ \frac{W_{3}}{3}, \;
\bar{W}=\frac{\bar{W}_{2}}{2}+ \frac{\bar{W}_{3}}{3},
\label{w23}
\end{eqnarray}
$W_{2}(\bar{W}_2)$ having two chiral superfields and $W_{3}(\bar{W}_3)$ 
three superfields. Explicitly we have that the term with two superfields is
given by: 
\begin{eqnarray}
W_{2}&=&\sum_{a=e}^{\tau}[\mu_{0a}\hat{L} \hat{ \eta}^{\prime}+
\mu_{1a}\hat{L} \hat{ \chi}^{\prime}]+ 
\mu_{ \eta} \hat{ \eta} \hat{ \eta}^{\prime}+
\mu_{ \chi} \hat{ \chi} \hat{ \chi}^{\prime}\nonumber \\ &+&
 \mu_{ \rho} \hat{ \rho} \hat{ \rho}^{\prime},
 \label{w2}
\end{eqnarray}
where $\hat{L} \hat{ \eta}^{\prime} \equiv \hat{L}_{i} 
\hat{ \eta}^{\prime}_{i}$. The term with three superfields is written explicitly
as 
\begin{eqnarray}
W_{3}&=& \sum_{a=e}^{\tau}\sum_{b=e}^{\tau}\lambda_{1ab} \hat{L}_{aL} 
\hat{ \rho}^{\prime} \hat{l}^{c}_{bL}+  
\sum_{a=e}^\tau[\lambda_{2a} \epsilon \hat{L}_{aL} \hat{\chi} \hat{\rho}+
\lambda_{3a} \epsilon \hat{L}_{aL} \hat{\eta} \hat{\rho}]+
\sum_{a=e}^\tau\sum^\tau_{b=e}\lambda_{4ab} \epsilon \hat{L}_{aL} \hat{L}_{bL} 
\hat{\rho} 
\nonumber \\ &+& 
\sum_{i=1}^{3}\kappa_{1i} \hat{Q}_{3L} \hat{\eta}^{\prime} \hat{u}^{c}_{iL}+
\kappa_{1}^{\prime} \hat{Q}_{3L} \hat{\eta}^{\prime} \hat{u}^{\prime c}_{L}+
\sum_{i=1}^{3}\kappa_{2i} \hat{Q}_{3L} \hat{\chi}^{\prime} \hat{u}^{c}_{iL}+
\kappa_{2}^{\prime} \hat{Q}_{3L} \hat{\chi}^{\prime} \hat{u}^{\prime c}_L+
\sum_{i=1}^{3}\kappa_{3i}\hat{Q}_{3L} \hat{\rho}^{\prime} \hat{d}^{c}_{iL}
\nonumber \\ &+&
\sum_{\alpha =1}^{2}\kappa^\prime_{3 \alpha}\hat{Q}_{3L} 
\hat{\rho}^{\prime} \hat{d}^{c}_{\alpha L} +
\sum_{\alpha=1}^{2}\sum_{i=1}^{3}\kappa_{3\alpha i} \hat{Q}_{\alpha L}
 \hat{\eta} \hat{d}^{c}_{iL}+
\sum_{\alpha=1}^{2}\sum_{\beta=1}^{2}\kappa^\prime_{3\alpha \beta}
\hat{Q}_{\alpha L} \hat{\eta} \hat{d}^{\prime c}_{\beta L}+
\sum_{\alpha =1}^{2}\sum_{i=1}^{3}\kappa_{4\alpha i} 
\hat{Q}_{\alpha L}\hat{\rho}\hat{u}^{c}_{iL} 
\nonumber \\ &+&
\sum_{\alpha =1}^{2}\kappa_{4\alpha}^{\prime} 
\hat{Q}_{\alpha L}\hat{\rho}\hat{u}^{\prime c}_{L}+
\sum_{\alpha=1}^{2}\sum_{i=1}^{3}\kappa_{5\alpha i} \hat{Q}_{\alpha L}
 \hat{\chi} \hat{d}^{c}_{iL} +
\sum_{\alpha=1}^{2}\sum_{\beta=1}^{2}\kappa^\prime_{5\alpha \beta} 
\hat{Q}_{\alpha L} \hat{\chi} \hat{d}^{\prime c}_{\beta L}
\nonumber \\ &+&
\sum_{\alpha =1}^{2} \sum_{\beta =1}^{2} \sum_{\gamma =1}^{2} 
\epsilon f_{1\alpha\beta\gamma}\hat{Q}_{\alpha L} \hat{Q}_{\beta L}
\hat{Q}_{\gamma L} +  
f_{2} \epsilon \hat{ \rho} \hat{ \chi} \hat{ \eta}+
f^{\prime}_{2} \epsilon \hat{ \rho}^{\prime}\hat{ \chi}^{\prime}
\hat{ \eta}^{\prime} +
\sum_{i=1}^{3}\sum_{\beta =1}^{2}\sum_{j=1}^{3}\xi_{1i \beta j} 
\hat{d}^{c}_{iL} \hat{d}^{\prime c}_{\beta L} \hat{u}^{c}_{j L}
\nonumber \\ &+&
\sum_{i=1}^{3}\sum_{\beta =1}^{2}\xi_{2i \beta } 
\hat{d}^{c}_{i L} \hat{d}^{\prime c}_{\beta L} \hat{u}^{\prime c}_{L}+
\sum_{i=1}^{3}\sum_{j=1}^{3}\sum_{j=1}^{3}\xi_{3ijk} 
\hat{d}^{c}_{iL} \hat{d}^{c}_{jL} \hat{u}^{c}_{k L} +
\sum_{i=1}^{3}\sum_{j=1}^{3}\xi_{4ij} \hat{d}^{c}_{i L} 
\hat{d}^{c}_{jL} \hat{u}^{\prime c}_{L} \nonumber \\ &+&
\sum_{\alpha =1}^{2}\sum_{\beta =1}^{2}\sum_{i=1}^{3}
\xi_{5 \alpha \beta i} \hat{d}^{\prime c}_{\alpha L}
\hat{d}^{\prime c}_{\beta L} \hat{u}^{c}_{iL}+
\sum_{\alpha =1}^{2}\sum_{\beta =1}^{2}\xi_{6 \alpha \beta} 
\hat{d}^{\prime c}_{\alpha L}\hat{d}^{\prime c}_{\beta L}
\hat{u}^{\prime c}_{L} + 
\sum_{a=e}^{\tau}\sum_{\alpha =1}^{2}\sum_{i=1}^{3}
\xi_{7a \alpha j}\hat{L}_{aL} \hat{Q}_{\alpha L} \hat{d}^{c}_{jL}
\nonumber \\ & +&
\sum_{a=e}^{\tau}\sum_{\alpha =1}^{2}\sum_{\beta =1}^{2}
\xi_{8a \alpha \beta}\hat{L}_{aL} \hat{Q}_{\alpha L} 
\hat{d}^{\prime c}_{\beta L}. 
\label{w3}
\end{eqnarray}

As usual it is necessary to introduce the so called soft terms that break the
supersymmetry explicitly.

\subsection{The soft terms}
\label{subsec:softterm}

 The soft terms  can be written as  
\begin{equation}
{\cal L}_{\mbox{soft}}={\cal L}_{GMT}+
{\cal L}^{\mbox{soft}}_{\mbox{Scalar}}+{\cal L}_{SMT},
\label{soft}
\end{equation}
where
\begin{eqnarray}
{\cal L}_{GMT}&=&- \frac{1}{2} [m_{ \lambda_{C}} \sum_{b=1}^{8} 
\left( \lambda^{b}_{C} \lambda^{b}_{C} \right) +m_{ \lambda} \sum_{b=1}^{8} 
\left( \lambda^{b}_{A} \lambda^{b}_{A} \right) \nonumber \\ 
&+&  m^{ \prime} \lambda_{B} \lambda_{B}+H.c.] ,
\label{gmt}
\end{eqnarray}
give mass to the boson superpartners and
\begin{eqnarray}
{\cal L}^{\mbox{soft}}_{\mbox{Scalar}}&=&
-m^2_{ \eta}\eta^{ \dagger}\eta-m^2_{ \rho}\rho^{ \dagger}\rho-
m^2_{ \chi}\chi^{ \dagger}\chi - 
m^2_{\eta^{\prime}}\eta^{\prime \dagger}\eta^{\prime}-
m^2_{\rho^{\prime}}\rho^{\prime \dagger}\rho^{\prime}-
m^2_{\chi^{\prime}}\chi^{\prime \dagger}\chi^{\prime}
\nonumber \\
&+&[k_1\epsilon_{ijk}\rho_i\chi_j\eta_k+ 
k^{\prime}_1\epsilon_{ijk}\rho^{\prime}_i\chi^{\prime}_j\eta^{\prime}_k+H.C.],
\label{potencial}
\end{eqnarray}
in order to give mass to the scalars, we have omitting the sum upon repeated
indices, $i,j,k=1,2,3$ . 
Finally, we have to add
\begin{eqnarray}
-{\cal L}_{SMT}&=& m^{2}_{L_a} \tilde{L}^{\dagger}_{aL}
\tilde{L}_{aL}+
 m_{la}^{2}\tilde{l}^{c \dagger}_{aL} \tilde{l}^{c}_{aL}+ 
m^2_{Q_3} 
\tilde{Q}^{\dagger}_{3L} \tilde{Q}_{3L}+
m_{Q_{\alpha L}}^{2} 
\tilde{Q}^{\dagger}_{\alpha L} \tilde{Q}_{\alpha L}+ m_{u_{i}}^2 
\tilde{u}^{c \dagger}_{iL} \tilde{u}^{c}_{iL}+ m_{d_{i}}^2 
\tilde{d}^{c \dagger}_{iL} \tilde{d}^{c}_{iL} \nonumber \\
&+&m_{u'}^{2} \tilde{u}^{\prime c \dagger}_L \tilde{u}^{\prime c}_L 
+ m_{d_{ \alpha}^{'}}^{2} 
\tilde{d}^{\prime c \dagger}_{ \beta L} \tilde{d}^{\prime c}_{ \beta L} +
[M_{a}^2 \tilde{L}_{aL} \eta^{\prime \dagger}+
M^{\prime 2}_a \tilde{L}_{aL} \chi^{\prime\dagger} +
\varepsilon_{1a} \tilde{L}_{aL} \rho^{\prime} \tilde{l}^{c}_{L}+ 
\varepsilon_{2a} \epsilon \tilde{L}_{aL} \chi \rho \nonumber \\ &+&
\varepsilon_{3a} \epsilon \tilde{L}_{aL} \eta \rho +
 \varepsilon_{4ab} \epsilon \tilde{L}_{aL} \tilde{L}_{bL} \rho 
+   
\varrho_{1i} \tilde{Q}_{3L} \eta^{\prime} \tilde{u}^{c}_{iL}+
\varrho_{1}^{\prime} \tilde{Q}_{3L} \eta^{\prime} \tilde{u}^{\prime c}_{L} + 
\varrho_{2i} \tilde{Q}_{3L} \chi^{\prime} \tilde{u}^{c}_{iL}+
\varrho_{2}^{\prime} \tilde{Q}_{3L} \chi^{\prime} \tilde{u}^{\prime c}_{L} 
 \nonumber \\
&+& 
\varrho_{3\alpha i} \tilde{Q}_{\alpha L} \eta \tilde{d}^{c}_{iL}+
\varrho_{3\alpha \beta}^{\prime} \tilde{Q}_{\alpha L} \eta 
\tilde{d}^{\prime c}_{\beta L}+
\varrho_{4\alpha i} \tilde{Q}_{\alpha L}\rho\tilde{u}^{c}_{iL} +
\varrho_{4\alpha}^{\prime} \tilde{Q}_{\alpha L}\rho
\tilde{u}^{\prime c}_{L}+\varrho_{5i}\tilde{Q}_{3L} \rho^{\prime}
 \tilde{d}^{c}_{iL}\nonumber \\ &+&
\varrho_{5 \alpha}^{\prime} \tilde{Q}_{3L} \rho^{\prime} 
\tilde{d}^{c}_{\alpha L} + 
\varrho_{6\alpha i} \tilde{Q}_{\alpha L} \chi \tilde{d}^{c}_{iL}+ 
\varrho_{6\alpha \beta}^{\prime} \tilde{Q}_{\alpha L} \chi 
\tilde{d}^{\prime c}_{\beta L}+
\varrho_{7\alpha\beta\gamma} \tilde{Q}_{\alpha L} \tilde{Q}_{\beta L} 
\tilde{Q}_{\gamma L}+ 
\upsilon_{1i \beta j} \tilde{d}^{c}_{iL} \tilde{d}^{\prime c}_{\beta L} 
\tilde{u}^{c}_{jL}\nonumber \\ &+&
\upsilon_{2i \beta } \tilde{d}^{c}_{iL} \tilde{d}^{\prime c}_{\beta L} 
\tilde{u}^{\prime c}_{L}+
\upsilon_{3ijk} \tilde{d}^{c}_{iL} \tilde{d}^{c}_{jL} \tilde{u}^{c}_{kL} 
+  
\upsilon_{4ij} \tilde{d}^{c}_{iL} \tilde{d}^{c}_{jL} \tilde{u}^{\prime c}_{L}
+
\upsilon_{5 \alpha \beta i} \tilde{d}^{\prime c}_{\alpha L}
\tilde{d}^{\prime c}_{\beta L} \tilde{u}^{c}_{iL}+
\upsilon_{6 \alpha \beta} \tilde{d}^{\prime c}_{\alpha L}
\tilde{d}^{\prime c}_{\beta L}
\tilde{u}^{\prime c}_{L}\nonumber \\ &+&   
\upsilon_{7a \alpha j}\tilde{L}_{aL} \tilde{Q}_{\alpha L} \tilde{d}^{c}_{jL} 
+
\upsilon_{8a \alpha \beta}\tilde{L}_{aL} \tilde{Q}_{\alpha L} 
\tilde{d}^{\prime c}_{\beta L}+H.C.], 
\label{xxx}
\end{eqnarray}
in order to give appropriate masses to the sfermions.

\section{The scalar Potential}
\label{sec:potential}

In the present model the scalar potential is written as
\begin{equation}
V_{331}=V_F+V_D+V_{\rm soft},
\label{p1}
\end{equation}
where
\begin{eqnarray}
V_F&=&-{\cal L}_F=\sum_m F^\dagger_m F_m 
\nonumber \\ &=&\sum_{ijk}[\left\vert\frac{\mu_\eta}{2}\eta^\prime_i+
\frac{f_2}{3}\epsilon_{ijk}\rho_j\chi_k
\right\vert^2+
\left\vert
\frac{\mu_\chi}{2}\chi^\prime_i+\frac{f_2}{3}\epsilon_{ijk}\eta_j\rho_k
\right\vert^2+
\left\vert
\frac{\mu_\rho}{2}\rho^\prime_i+\frac{f_2}{3}\epsilon_{ijk}\chi_j\eta_k
\right\vert^2\nonumber \\ &+&
\left\vert
\frac{\mu_\eta}{2}\eta_i+\frac{f^\prime_2}{3}\epsilon_{ijk}\rho^\prime_j
\chi^\prime_k\right\vert^2+
\left\vert
\frac{\mu_\chi}{2}\chi_i+\frac{f^\prime_2}{3}\epsilon_{ijk}
\eta^\prime_j\rho^\prime_k\right\vert^2+
\left\vert \frac{\mu_\rho}{2}\rho_i+\frac{f^\prime_2}{3}\epsilon_{ijk}
\chi^\prime_j\eta^\prime_k\right\vert^2]
\label{p2}
\end{eqnarray}
and
\begin{eqnarray}
V_D&=&-{\cal
L}_D=\frac{1}{2}(D^aD^a+DD)=\frac{g^{\prime2}}{18}(-\eta^\dagger\eta+
\eta^{\prime\dagger}\eta^\prime-\chi^\dagger\chi+\chi^{\prime\dagger}\chi^\prime
+2\rho^\dagger\rho-2\rho^{\prime\dagger}\rho^\prime)^2\nonumber \\ &+&
\frac{g^2}{8}(\eta^\dagger_i\lambda^b_{ij}\eta_j-
\eta^{\prime\dagger}_i\lambda^{*b}_{ij}\eta^\prime_j+
\chi^\dagger_i\lambda^b_{ij}\chi_j-
\chi^{\prime\dagger}_i\lambda^{*b}_{ij}\chi^\prime_j+
\rho^\dagger_i\lambda^b_{ij}\rho_j+
\rho^{\prime\dagger}_i\lambda^{*b}_{ij}\rho^\prime_j)^2\!,
\label{p3}
\end{eqnarray}
finally,
\begin{eqnarray}
V_{\rm soft}&=&-{\cal L}_{\rm soft}= 
m^2_\eta\eta^\dagger\eta+m^2_\rho\rho^\dagger\rho+m^2_\chi\chi^\dagger\chi
+m^2_{\eta^\prime}\eta^{\prime\dagger}\eta^\prime \nonumber \\ &+&
m^2_{\rho^\prime}\rho^{\prime\dagger}\rho^\prime+
m^2_{\chi^\prime}\chi^{\prime\dagger}\chi^\prime-
\epsilon_{ijk}(k_1 \rho_i\chi_j\eta_k+
k^\prime_1\rho^\prime_i\chi^\prime_j\eta^\prime_k\nonumber \\&+&H.c.),
\label{p4}
\end{eqnarray}
where we have used the scalar multiplets in Eqs.~(\ref{esctrip}) and
(\ref{escac}). 

With Eqs.~(\ref{p2})-(\ref{p4}) we can work out the mass spectra of the scalar
and pseudoscalar fields by making the usual shift
$X^0\to\frac{1}{\sqrt2}(v_X+H_X+iF_X)$. The analysis is similar to that of
Ref.~\cite{331susy} and we will not write the constraints equation, etc.

By using as input the following values for the parameters:
$\sin^2\theta_W=0.2314$, $g=0.6532$, $g^\prime=1.1466$; $f_2=2$,
$f^\prime_2=10^{-3}$; 
$k_1=k^\prime_1=10$ GeV; $\mu_\eta=\mu_\rho=\mu_\chi=-10^3$ GeV; 
$m_\eta=15$ GeV, $m_\rho=10$ GeV. $m_\rho=244.99$ GeV;
$m_{\chi_2}=m_{\chi^\prime_2}=10^3$ GeV and $m_{\rho^\prime}=13$ GeV, we obtain
the  masses
\begin{eqnarray}
&&m_1\approx1702, m_2\approx1449, m_3\approx387, \nonumber \\
&&m_4\approx380,m_5\approx361, m_6\approx130,
\label{mi}
\end{eqnarray}
for the scalar sector (all masses are in GeV). Note that the lightest neutral
scalar is heavier than the lower limit of the Higgs scalar of the standard
model, i.e., $m_H\stackrel{>}{\sim} 114$ GeV.  
For the pseudoscalar sector we obtain
\begin{eqnarray}
&&M_1\approx1702, M_2\approx1449,M_3\approx363,\nonumber \\
 &&M_4\approx5, M_5=0, M_6=0,
\label{Mi}
\end{eqnarray}
only the two massless pseudoscalars are exact values, i.e.,
there are two Goldstone bosons as it should be. Notice that there is a light
pseudoscalar. Although the values of the masses above are only an exercise, 
and it is possible that other values of the input parameters would give another 
set of masses, we would like to point out the following. In the basis 
$(F_{\eta_1}, F_\rho, F_{\chi_2},F_{\eta^\prime_1},F_{\rho^\prime},
F_{\chi^\prime_2})$ the
lightest pseudoscalar is given by $(0.0099,0.0012,0.7070,0.0071,0.0170,0.7070)$,
hence we see that it is mainly $F_{\chi_2}$ and $F_{\chi^\prime_2}$. So, we need
to investigate the couplings of these pseudoscalars with the $Z^0$.

In the  ($W_3,W_8,B$) basis we have the mass square of the real vector bosons 
given by:
\begin{equation}
\frac{g^2}{4}\,\left(
\begin{array}{ccc}
V^2+U^2 & \frac{1}{\sqrt3}\left(V^2-U^2 \right)
&-\frac{2t}{3}\left(V^2+2U^2\right) \\
 & \frac{1}{3}\left(V^2+U^2+4W^2 \right) &
 -\frac{2t}{3\sqrt3}\left(V^2-2U^2-2W^2 \right) \\
 & & \frac{4t^2}{9}\left( V^2+4U^2+W^2\right)\!,
\end{array}
\right)
\label{massneutros}
\end{equation} 
where we have defined $V^2=v^2+v^{\prime 2}$, $U^2=u^2+u^{\prime 2}$
and $W^2=w^2+w^{\prime 2}$, and
\begin{equation}
t^2=\left(\frac{g^\prime}{g}\right)^2=\frac{\sin^2\theta_W}{1-\frac{4}{3}
\sin^2\theta_W}. 
\label{sin331}
\end{equation}

The eigenstates of Eq.~(\ref{massneutros}) are 
\begin{equation}
A=\frac{\sqrt3}{4t^2+3}\left(t\,W_3-\frac{t}{\sqrt3}W_8+B \right),
\label{foton}
\end{equation}
for the photon, and 
\begin{equation}
Z^0\approx\frac{3t}{4t^2+15t^2+9}\left[ -\left(\frac{t^2+3}{3t} 
\right)W_3-\frac{t}{\sqrt3}W_8+B\right],
\label{z}
\end{equation}
and
\begin{equation}
Z^{0\prime}\approx\frac{t}{t^2+3}\left( \frac{\sqrt3}{t}W_8+B\right)
\label{zp}
\end{equation}
for the $Z^0$ and $Z^{0\prime}$, we have neglected the mixing among $Z^0$ and
$Z^{0\prime}$, so that $M^2_Z/M^2_W\approx(3+4t^2)/(3+t^2)=1/\cos^2\theta_W$.
With this at hand, we were able to verify that the couplings of
$F_{\chi_2},F_{\chi^\prime_2}$ are given by the usual vertex of the Higgs scalar
in the standard model times a factor proportional to $(W/v_W)(U/W)^4$ or
$(W/v_W)(V/W)^4$, where $v_W\approx246$ GeV, and these couplings go to zero as
$W\to \infty$. This behavior is expected since both
$\chi^0_{2},\chi^{0\prime}_2$ are singlets of $SU(2)_L\otimes U(1)_Y$ and do
not couple to the $Z^0$ in this limit.   

For completeness we show that the lightest scalar, in the basis 
$(H_{\eta_1}, H_\rho,H_{\chi_2},H_{\eta^\prime_1},H_{\rho´},H_{\chi^\prime_2})$,
is written as $(-0.0581,-0.9775,0.0610,-0.0394,-0.0592,0.1800)$, that is, 
it is mainly
$H_\rho$ which transforms as doublet under $SU(2)_L\otimes U(1)_Y$. 

\section{Conclusions}
\label{sec:con}

We have put forward a supersymmetric version of the 3-3-1 model of
Refs.~\cite{mpp} which includes right-handed neutrinos transforming
non-trivially under $SU(3)_L$. This sort of models is an interesting possibility
since neutrinos are massive particles and right-handed neutrinos should
to be added eventually to any extension of the standard model.

Concerning the scalar and pseudoscalar mass spectra we have found two different
situations: for the scalar sector we were able to find all the Higgs scalars
with masses above the $Z^0$ mass and above the lower limit of the standard model
Higgs boson obtained by LEP: $m_H\stackrel{>}{\sim}114$ GeV; for the
pseudoscalars, for the same set of the input parameters, we have found a
considerably light one ($M_4=5$ GeV) which in principle can bring some 
problems. However, a carefully analysis have shown that the mass eigenstate
corresponding to $M_4$ is mainly formed by the symmetry eigenstates $F_{\chi_2}$
and $F_{\chi^\prime_2}$, and studying the couplings of these pseudoscalars with
the $Z^0$ we have noted that they vanish in the limit where $w$ and $w^\prime$
go to infinity {\it i.e.}, both pseudoscalars decouple from $Z^0$ in this limit
since they are singlets under $SU(3)_L\otimes U(1)_Y$.  

In the other supersymmetric 3-3-1 model~\cite{331susy}, the proton decay modes
are $p\to K^+\bar{\nu}_\mu$ and $p\to K^+e^\mp \mu^\pm\bar{\nu}_\tau$ but in the
present one, only the mode $p\to \pi^+\bar{\nu}_\mu$ is possible. It means that
the constraints coming from SuperKamiokande data on $p\to K^+\bar{\nu}_e$, 
which implies that $\tau_P>10^{33}$ years in this decay mode~\cite{sk99}, are
avoided. 

However, there are higher-dimension (nonrenormalizable) operators, 
that arise from new physics at some scale $\Lambda$. For instance, there are 
dimension-5 operators that violate baryon or lepton number, that are 
allowed by the gauge invariance, and that contribute to the
proton decay unless they are sufficiently suppressed. 
In the context of the MSSM we have~\cite{d5}
\begin{eqnarray}
&&\frac{\kappa^{1}_{ijkl}}{\Lambda}\hat{Q}_{i}\hat{Q}_{j}\hat{Q}_{k}\hat{L}_{l}
+ 
\frac{\kappa^{2}_{ijka}}{\Lambda}\hat{u}^{c}_{i}
\hat{u}^{c}_{j}\hat{d}^{c}_{k}\hat{l}^{c}_{a}.
\label{genial}
\end{eqnarray}
These terms contribute to the proton decay with diagrams at 1-loop 
level involving gauginos and gluinos, known as
dressing diagrams. These are the dangerous terms that induce the decay mode $p 
\rightarrow K^{+} \bar{\nu}_e$.
This channel is enough to exclude the minimal supersymmetric SU(5)
model~\cite{susysu5} since the SuperKamiokande data~\cite{sk99}.  

In the present model dimension-5 operators that violate lepton and baryon
number, that are allowed by the 3-3-1 symmetry, may arise
in unification theories~\cite{ponce}, and are given by
\begin{eqnarray}
&&\frac{\kappa^{1}_{a \alpha 
\beta}}{\Lambda}(\hat{L}_{a}\hat{Q}_{\alpha})(\hat{Q}_{3}\hat{Q}_{\beta})+
\frac{\kappa^{2}_{aijk}}{\Lambda}\hat{l}^{c}_{a}\hat{u}^{c}_{i}
\hat{u}^{c}_{j}\hat{d}^{c}_{k}+
\frac{\kappa^{3}_{abi}}{\Lambda}(\hat{L}_{a}\hat{L}_{b})
(\hat{Q}_{3}\hat{d}^{c}_{i})+
\frac{\kappa^{4}_{\alpha ij}}
{\Lambda}(\hat{Q}_{\alpha}\hat{Q}_{3})\hat{u}^{c}_{i}\hat{d}^{c}_{j} 
\nonumber \\ &+&
\frac{\kappa^{5}_{ab}}{\Lambda}(\hat{L}_{a}\hat{\eta}^{\prime})(\hat{L}_{b}
\hat{\eta}^{\prime})+
\frac{\kappa^{6}_{ab}}{\Lambda}(\hat{L}_{a}\hat{\chi}^{\prime})
(\hat{L}_{b}\hat{\chi}^{\prime})+
\frac{\kappa^{7}_{\alpha \beta}}
{\Lambda}(\hat{Q}_{3}\hat{Q}_{\alpha})(\hat{Q}_{\beta}\hat{\eta})+
\frac{\kappa^{8}_{\alpha 
\beta}}{\Lambda}(\hat{Q}_{3}\hat{Q}_{\alpha})(\hat{Q}_{\beta}\hat{\chi}),
\label{genial2}
\end{eqnarray}
while in the SUSY 3-3-1 model of Ref.~\cite{331susy} these operators are
\begin{eqnarray}
&&\frac{\kappa^{1}_{a \alpha 
\beta}}{\Lambda}(\hat{L}_{a}\hat{Q}_{\alpha})(\hat{Q}_{3}\hat{Q}_{\beta})+
\frac{\kappa^{2}_{\alpha 
\beta}}{\Lambda}(\hat{Q}_{\alpha}\hat{J}^{c})(\hat{Q}_{3}\hat{j}^{c}_{\beta})+
\frac{\kappa^{3}_{abi}}{\Lambda}(\hat{L}_{a}\hat{L}_{b})(\hat{Q}_{3}
\hat{u}^{c}_{i})\nonumber \\ &+&
\frac{\kappa^{4}_{\alpha 
\beta}}{\Lambda}(\hat{\eta}\hat{Q}_{\alpha})(\hat{Q}_{3}\hat{Q}_{\beta}) 
+
\frac{\kappa^{5}_{ab}}{\Lambda}(\hat{L}_{a}\hat{\eta}^{\prime})(\hat{L}_{b}
\hat{\eta}^{\prime}).
\label{genial1}
\end{eqnarray}
The suppression of the effective operators in Eqs.~(\ref{genial2}) and
(\ref{genial1}) in the context of SUSY 3-3-1 models will be considered
elsewhere~\cite{marcos2}. 

Finally, we would like to say that the 3-3-1 model with non-supersymmetric
right-handed neutrinos furnishes a good candidate for self-interacting dark 
matter (SIDM) since there are two Higgs bosons, one 
scalar, and one pseudoscalar, which have the properties of the candidates for
dark matter like, stability, neutrality, and that they must not overpopulate the
universe~\cite{longlan} and, in particular for a self-interacting dark matter
candidates, they have large scattering cross-section and negligible annihilation
or dissipation. It means that the candidate for SIDM has not to be introduced
{\it ad hoc} as in other models~\cite{sidm}. This feature remains valid in the
supersymmetric version that we have developed in this work, but it deserves a
more careful study. 

\acknowledgments 
This work was partially supported by CNPq under the processes
305185/2003-9 (JCM) and 306087/88-0 (VP).

\end{document}